\documentclass[aps,pra,twocolumn,nofootinbib,longbibliography,10pt]{revtex4}
\usepackage{hyperref}
\usepackage{comment}
\usepackage{graphicx} 
\usepackage{xcolor}
\usepackage{amsmath}
\usepackage{amsfonts}
\usepackage{amssymb}
\usepackage{graphicx}
\usepackage{dcolumn}
\usepackage{bm}
\usepackage{hyperref,etoolbox}
\usepackage{mathrsfs}
\usepackage{subfig}
\usepackage{natbib}
\usepackage{caption}
\usepackage[none]{hyphenat}
\newcommand{\ket}[1]{\left|#1\right>}
\newcommand{\bra}[1]{\left<#1\right|}
\newcommand{\braket}[2]{\left<#1\mid#2\right>}

\newcommand{\comgen}[2]{\left[#1,#2\right]}
\newcommand{\anticom}[2]{\left\{#1,#2\right\}}

\usepackage{lipsum}

\begin{document}
\title{Current and quantum transport factor of fermionic system in fermionic bath}

\author{\textbf{Jayarshi Bhattacharya}}
\email{dibyajayarshi@gmail.com}
\author{\textbf{Gautam Gangopadhyay}}
\email{gautam@bose.res.in}
\author{\textbf{Sunandan Gangopadhyay}}
\email{sunandan.gangopadhyay@gmail.com}
\affiliation{\textit{S.N. Bose National Centre for Basic Sciences, JD Block, Sector-III, Salt Lake, Kolkata 700106, India}}

\begin{abstract}
	\noindent This paper explores the dynamics of current and the quantum transport factor in a fermionic system with a central oscillator interacting with two fermionic reservoirs at different temperatures. We derive the master equation for the system’s density matrix, accounting for energy exchange between the system and the reservoirs. The current is analyzed in relation to system parameters and reservoir temperatures, revealing that the quantum transport factor differs from classical systems, approaching Carnot efficiency at high temperatures but being different at low temperatures. We also examine the power spectrum of the current, providing insights into current fluctuations and their temperature dependence. Furthermore, we derive the Fokker-Planck equation and the Glauber-Sudarshan P-representation, where the steady-state probability distribution takes a Gaussian form, confirming the system behaves like a harmonic oscillator. This work advances the understanding of quantum transport in fermionic systems and provides a foundation for future research in quantum thermodynamics.
\end{abstract}

\maketitle
\section{Introduction}
\noindent It has been a topic of interest for a long time as we see that open quantum systems and its thermodynamic properties are studied in \cite{abe2011similarity,gemmer2009quantum,alicki2018introduction}. Since 1960s, people studied different quantum system and bath \cite{sudarshan1963equivalence,LachsPaper,Glauber1963,Cahill1969}. The harmonic oscillator or two level atomic system coupled with a single bath has been studied extensively in \cite{Carmichael,Zubairy,Louisell}. But since the publication of \cite{cahill1999density}, the fermionic oscillator in bosonic or fermionic bath has been studied as well. In recent time, the open quantum systems has been a hot topic of interest for computation, technology and understanding  \cite{weinbub2023computational, shaker2023advancements}.\\
Physicists have always been interested in systems connected to thermal baths \cite{Mukherjee:2024egj}. In this article, we focus on a fermionic system linked to two different fermionic baths at different temperatures. Previously, in \cite{karmakar2016fermionic}, a fermionic oscillator was studied, connected to an emitter (a fermionic bath at temperature $T_e$ and a collector (a fermionic bath at temperature $T_c$, where the coupling constants were Grassmann variables. However, other studies \cite{ghosh2012fermionic,moitra2020entanglement} showed that even when the coupling constants are ordinary numbers, meaningful results can still be obtained.\\
Here, we use the standard method to derive the master equation for the density operator under the Markovian approximation, as detailed in \cite{Carmichael,Zubairy,pearle2012simple}. On the other hand, non-Markovian systems have also been explored in works like \cite{PhysRevB.92.235440, PhysRevB.95.064308, PhysRevD.45.2843}, which analyze various open quantum systems to derive exact master equations. For instance, \cite{PhysRevB.92.235440} examined a single oscillator coupled to an electron bath to calculate heat flow. In this article, however, we focus solely on Markovian systems to obtain useful insights.\\
In this work, we first derived the master equation for a fermionic system connected to a fermionic emitter and a fermionic collector. This result provided the foundation for discussing the particle density under flux balance conditions. From the flux balance, we applied the method from \cite{bhattacharya2025current} to calculate the current flowing through the system, from the emitter to the collector. This current can be thought of as tracking the flow of water through a series of pipes. In the next section, we introduced a quantum transport factor to describe how smoothly the current flows, without leaks or blockages.\\
It is shown in \cite{bhattacharya2025current} that the quantum transport factor has similarities with energy efficiency. Advancements in the energy transport factor have become a key focus in both classical and quantum thermodynamics, attracting significant interest from the scientific community. These developments have led to major technological progress \cite{cakmak2020construction, PhysRevE.76.031105, unknown2023comparative}. We focused our work on quantum transport factor such a way that it gives us idea about efficiency of the system. We found that the quantum transport factor is mathematically identical to the Carnot efficiency \cite{PhysRevLett.134.027101, fermi2012thermodynamics} when the emitter and collector temperature goes to infinity. Earlier in \cite{rezek2006irreversible, Mukherjee:2022prn, wang2009performance}, it is found that the quantum efficiency is equal to the classical counter parts. In general, studies \cite{moreira2020enhancing,li2022quantum} have shown that the efficiency is higher in non-Markovian systems compared to Markovian systems. But in our analysis we found that the quantum transport differs from the classical Carnot efficiency even under Markov approximation. While the bosonic case \cite{bhattacharya2025current} shows that the quantum transport factor is always greater than the Carnot efficiency, the fermionic transport factor is revealed to be lesser than the Carnot efficiency at very low temperature.\\
Later, we analyzed the power spectrum of the current, as introduced in Section \ref{sec1}. To do this, we followed the standard procedure outlined in \cite{Carmichael} to calculate the current-current correlation, which allowed us to define the power spectrum \cite{huang2009introduction}. This helped us understand how the power spectrum changes with temperature and provided deeper insights into the system’s behavior. Interestingly, our findings show a qualitative agreement with the results reported in \cite{karmakar2016fermionic}.\\
Lastly we derived the Fokker-Planck equation from the master equation derived earlier. This helps us to understand the probabilistic distribution of the particles over time. But for fermionic operators we needed to introduce Grassmann algebra as shown in \cite{cahill1999density}. In recent years, many studies have explored fermionic systems using Grassmann algebra. For instance, the non-Markovian dynamics of a two-level atom in an electromagnetic field bath has been analyzed in \cite{PhysRevA.62.033821} by representing the atom with Grassmann variables and using coherent states for the field to express the transition amplitude. Additionally, interacting fermionic dynamics has been investigated in \cite{PhysRevA.64.025801} through stochastic linear transformations of Grassmann algebra and Bargmann states, particularly in studying Cooper-like pairing in trapped one-dimensional fermions. Another notable advancement is the extension of response function formulations from bosons to fermions by incorporating Grassmann sources in the Hamiltonian as shown in \cite{plimak2009causal}.\\
In this paper, we approach the problem of fermionic transport by utilizing the mathematical parallel between bosonic and fermionic fields within the framework of Glauber coherent states for fermions. This enables us to develop an algebraic method using Grassmann variables to compute expectation values of fermionic operators, avoiding the complexity of handling $2^{2n}$ c-numbers for $n$-fermionic states as done in \cite{plimak2009causal}, which can be challenging even in simplified models. We also demonstrate that the Fokker-Planck equation for the $P$ distribution in Grassmann variables qualitatively retains the same structure as its bosonic counterpart. In our work we have advanced our work for the non-steady state as well. For that reason, we find the time dependent solution of the $P$ distribution. At steady state, the result matches previous results found in \cite{ghosh2012fermionic, cahill1999density}.\\
This article is structured as follows. In Section \ref{sec1}, we introduce the concept of current, which then leads to the definition of the quantum transport factor in Section \ref{sec2}. In Section \ref{sec3} we discussed about the power spectrum of the current. Section \ref{sec4} focuses on the Fokker-Planck equation for our system. Finally, we present our conclusions in Section \ref{sec5}.

\section{Current for Fermions}\label{sec1}
\noindent In this article, we analyze a fermionic oscillator that interacts with two separate reservoirs, an emitter and a collector. Each of these reservoirs is maintained at a different temperature, denoted by $T_e$ for the emitter and $T_c$ for the collector. What makes this setup particularly interesting is that both reservoirs are fermionic in nature, meaning they follow Fermi-Dirac statistics rather than classical or bosonic distributions.\\
To describe the dynamics of this system, we construct the Hamiltonian, which captures the interactions between the oscillator and the reservoirs. This Hamiltonian accounts for the internal energy of the fermionic oscillator, the energy of the emitter and collector, and the coupling between the oscillator and each reservoir. By carefully analyzing this Hamiltonian, we can study key properties such as energy transport, dissipation, and quantum correlations within the system. The Hamiltonian we start with reads
\begin{subequations}\label{hamiltonian}
	\begin{align}
		\hat{H_s}&=\hbar\omega_s \hat{a}^\dagger \hat{a}\\
		\hat{H_e}&=\hbar\sum\limits_{j}\omega^e_j \hat{b}_j^\dagger \hat{b}_j\label{1b}\\
		\hat{H_c}&=\hbar\sum\limits_{j}\omega^c_j \hat{d_j}^\dagger \hat{d_j}\label{1c}\\
		\hat{H_i}&=\hbar\sum\limits_{j}\left( \Gamma_{ej} \hat{a}^\dagger \hat{b}_j +\Gamma_{ej}^* \hat{a} \hat{b}^\dagger_j +\Gamma_{cj} \hat{a}^\dagger \hat{d_j} +\Gamma_{cj}^* \hat{a} \hat{d_j}^\dagger \right)
	\end{align}
\end{subequations}
In this context, $\hat{a}$, $\hat{b}_j$, and $\hat{d_j}$ represent operators satisfying the fermionic anti-commutation relations $\left\{\hat{b}_j,\hat{b_k}^\dagger\right\}=\delta_{jk}=\left\{\hat{d_j},\hat{d_k}^\dagger\right\}$ and $\left\{\hat{a},\hat{a}^\dagger\right\}=1$. The Hamiltonian described above, which models a single fermionic oscillator interacting with two reservoirs at distinct temperatures, holds significance in the study of quantum particle transport. Specifically, investigating transport behavior in systems with discrete energy levels has been a topic of considerable interest \cite{postma2001carbon,datta1997electronic,liang2002kondo}, making the model in this work particularly relevant. Earlier in \cite{karmakar2016fermionic}, similar model was considered, but with Grassmannian coupling. In real world, the couplings are usually meant to be found as ordinary numbers. So in this work we kept the coupling constants as ordinary numbers and found useful results.\\
The quantum current to be discussed later can be realized by configuring a source-system-sink setup, where the source and sink are bosonic reservoirs characterized by the Hamiltonians provided in eqs. \eqref{1b} and \eqref{1c}. Additionally, it is worth noting that the problem of a single Fermionic oscillator coupled to a reservoir also has significance in quantum optics. Here, the reservoir oscillators can correspond to the Fourier modes of the radiation field into which an excited atom undergoes spontaneous emission \cite{Carmichael}.\\
We employ the density matrix approach in the interaction picture to further our analysis. The differential equation \cite{Carmichael, Zubairy, Louisell} governs the time development of the density matrix $\hat{\rho_s}$ in the above picture.

\begin{align}\label{diffeq}
	\frac{d\hat{\rho_s}}{dt}&=-\int_0^{t}dt' ~~\text{Tr}_{e,c}\left[\hat{H_i}(t),\left[\hat{H_i}(t'),\hat{\rho}(t')\right]\right]~.
\end{align}
where, $\hat{\rho}(t')=\hat{\rho_s}(t')\otimes\hat{\rho_e}(0)\otimes\hat{\rho_c}(0)$.\\
With $\hat{H_i}(t)$ standing for the interaction Hamiltonian, this equation depicts the intricate relationship amongst the system and its surroundings. It is essential to understanding our fermionic system's dynamics.\\
\begin{widetext}
It is then possible to calculate the temporal evolution of the density matrix $\hat{\rho}$ in the Schr\"{o}edinger picture by applying the methods previously employed in \cite{Carmichael,Zubairy,karmakar2016fermionic}.
\begin{align}
	\frac{d\hat{\rho}}{dt}&=-i\omega\left[\hat{a}^\dagger \hat{a},\hat{\rho}\right]- \frac{\gamma_e+\gamma_c}{2}\left(\hat{a}^\dagger\hat{a}\hat{\rho}+\hat{\rho}\hat{a}^\dagger \hat{a}-2\hat{a}\hat{\rho}\hat{a}^\dagger\right)- \left(\gamma_e\bar{n}_e+\gamma_c\bar{n}_c\right)\left(\hat{\rho}-\hat{a}\hat{\rho}\hat{a}^\dagger-\hat{a}^\dagger\hat{\rho}\hat{a}\right)\label{masteq}
\end{align}
\end{widetext}
where $\omega=\omega_s+\Delta+\Delta'$ is the shifted frequency of the system.\\
The coefficients in question are essential for measuring the system's dissipative processes.  In addition, $\bar{n}_e$ and $\bar{n}_c$ represent the number densities, which are specified as
\begin{subequations}\label{n}
	\begin{align}
		\bar{n}_e&=\frac{1}{e^{\frac{\hbar\omega_s}{k_B T_e}}+1}\\
		\bar{n}_c&=\frac{1}{e^{\frac{\hbar\omega_s}{k_B T_c}}+1}~.
	\end{align}
\end{subequations}
These number densities reflect the equilibrium distributions of fermions within the emitter and collector reservoirs.\\
It is important to note that the master equation found in eq. \eqref{masteq} looks quite different to what found in \cite{karmakar2016fermionic} due to the coupling coefficient being ordinary numbers and not the Grassmanian ones. Nevertheless, we shall see that the current can be found in similar form as found in \cite{karmakar2016fermionic}.\\
To gain insight into the dynamics of this fermionic system, it is essential to analyse the average number of fermions within the system. Represented as $\left<\hat{n}(t)\right>$, this quantity can be obtained by studying the time evolution of the system's density matrix, yielding
\begin{align}\label{number}
	\left<\dot{\hat{n}}\right>(t) &=\left<\hat{a}^\dagger\hat{a}\dot{\hat{\rho}}(t)\right>\nonumber\\
	&=-\left(\gamma_e+\gamma_c\right)\left(\left<\hat{n}\right>(t)-\frac{\gamma_e\bar{n}_e+\gamma_c\bar{n}_c}{\gamma_e+\gamma_c}\right)~.
\end{align}
\begin{widetext}
    Solving this differential equation gives
    \begin{align}
    	\left<\hat{n}\right>(t)&= \frac{\gamma_e\bar{n}_e+\gamma_c\bar{n}_c}{\gamma_e+\gamma_c} +\left(\left<\hat{n}(0)\right>-\frac{\gamma_e\bar{n}_e+\gamma_c\bar{n}_c}{\gamma_e+\gamma_c}\right)e^{-(\gamma_e+\gamma_c)t}\label{extra1}~.
    \end{align}
\end{widetext}
This statement concisely illustrates the evolution of the average number of fermions over time, taking into consideration the damping effects controlled by $\gamma_e$ and $\gamma_c$ as well as the initial state of the system, represented as $\left<\hat{n}(0)\right>$.\\
To proceed, we analyse the steady-state number density, symbolised by $\bar{n}_s$, which sheds light on the system's long-term behavior. By setting $\left<\dot{\hat{n}}\right>(t)=0$ in eq. \eqref{number}, the steady-state value is obtained as  
\begin{align}  
	\bar{n}_s = \frac{\gamma_e\bar{n}_e + \gamma_c\bar{n}_c}{\gamma_e + \gamma_c}\label{steadynumber}~.  
\end{align}  
Interestingly, this result can also be reformulated as  
\begin{align}  
	\gamma_e(\bar{n}_e - \bar{n}_s) = \gamma_c(\bar{n}_s - \bar{n}_c)\label{balance}~.  
\end{align}  
Equation \eqref{steadynumber} expresses the steady-state number density, illustrating how the system stabilizes over time. Meanwhile, eq. \eqref{balance} highlights the equilibrium established between the emitter and collector reservoirs, with $\gamma_e$ and $\gamma_c$ playing critical roles. This equation reflects the flux balance achieved in steady-state conditions.\\
In this framework, we then explore the idea of current. $\frac{\gamma_e\gamma_c}{\gamma_e + \gamma_c}(\bar{n}_e - \bar{n}_c)$ is a rewrite of the left-hand side of eq. \eqref{balance}, $\gamma_e(\bar{n}_e - \bar{n}_s)$. This form demonstrates the inherent connection between the resulting quantum current and the imbalance in the emitter and collector reservoirs. Interestingly, a well-known result in the literature is the factor $\frac{\gamma_e\gamma_c}{\gamma_e + \gamma_c}$ \cite{karmakar2016fermionic, sun1999quantum}. The steady-state current is described by this relationship, which enables us to determine the average current of the system as follows
\begin{align}
	I(t)&=\frac{1}{2}\left[\gamma_e\left({\bar{n}_e}-\left<\hat{n}\right>(t)\right)+\gamma_c\left(\left<\hat{n}\right>(t)-{\bar{n}_c}\right)\right]\label{current}~.
\end{align}
Substituting $\left<\hat{n}\right>(t)$ from eq. \eqref{extra1} in the above equation leads to
\begin{align}
	I(t)&=I_s+\left(I_0-I_s\right) e^{-\left(\gamma_e+\gamma_c\right)t}\label{current_extra1}
\end{align}
where $I_0$ and $I_s$ are given by
\begin{align}
	I_0&=\frac{1}{2}\left[\gamma_e\left({\bar{n}_e}-\left<\hat{n}\right>(0)\right)+\gamma_c\left(\left<\hat{n}\right>(0)-{\bar{n}_c}\right)\right]~;\label{i0}\\
	I_s&= \frac{\gamma_e\gamma_c}{\gamma_e+\gamma_c}(\bar{n}_e-\bar{n}_c)\label{is}~.
\end{align}
When $\bar{n}_e=1$ and $\bar{n}_c=0$, this is perfectly in line with the relation found in \cite{karmakar2016fermionic}, which is $I_s=\frac{\gamma_e\gamma_c}{\gamma_e+\gamma_c}$. This demonstrates that the current has a comparable nature even when the coupling constants between the fermionic system and the fermionic bath are not Grassmanian as shown in \cite{karmakar2016fermionic}. This beaautiful correspondance can only be found because we defined the current differently in this work. Here the idea of flux balance is used which can be considered more of a physical consideration.
\section{quantum transport factor}\label{sec2}
\begin{figure}[ht]
    \centering
    \includegraphics[width=0.9\linewidth]{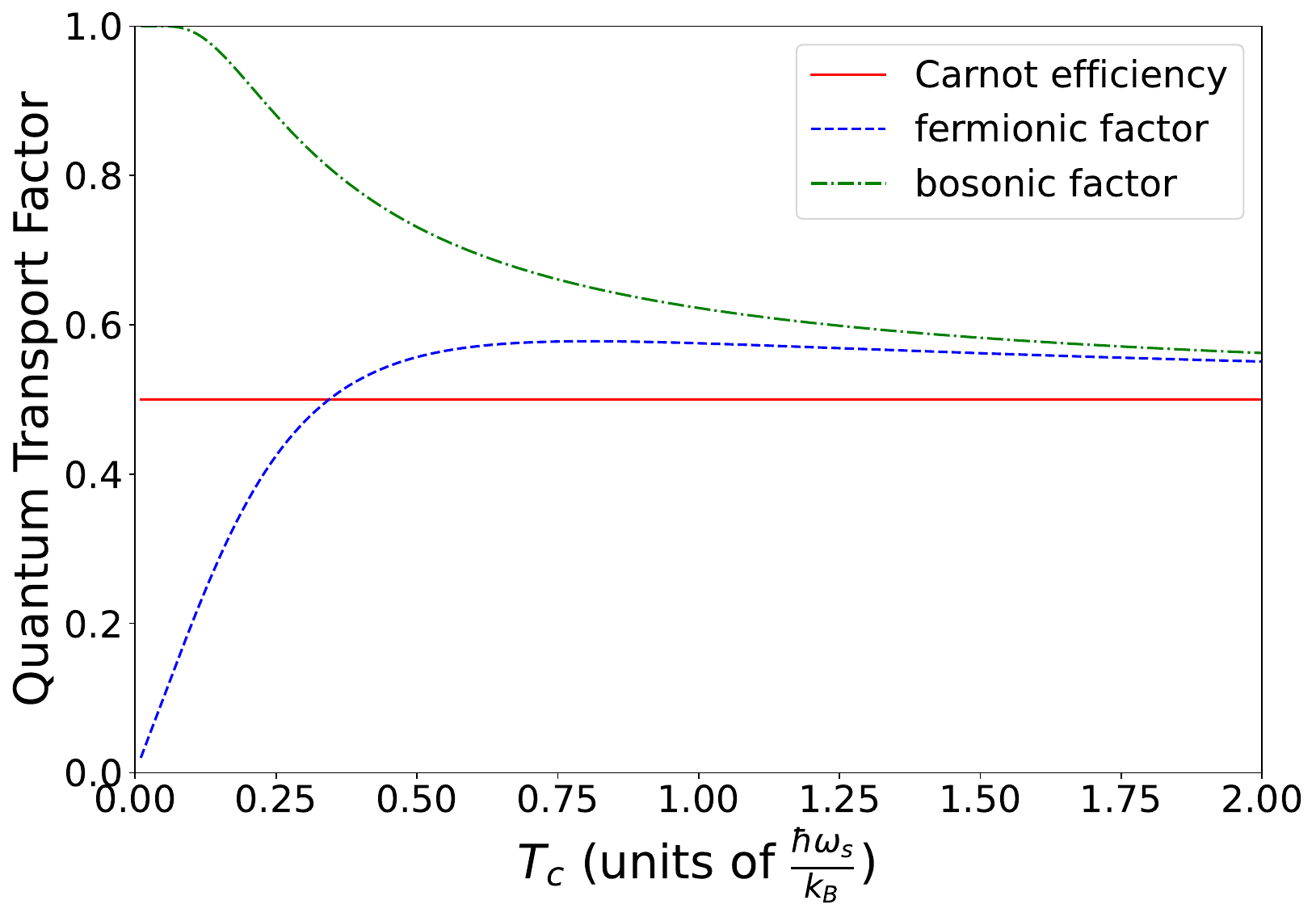}
    \caption{Quantum transport factor of fermionic system in fermionic bath, bosonic system in bosonic bath and classical Carnot efficiency is compared while $T_e=2T_c$ kept constant.}
    \label{fig:fig1}
\end{figure}

\noindent We now follow the technique used in \cite{bhattacharya2025current} to define quantum transport factor. We introduce the concept of steady-state energy loss per unit time, denoted as $E_s$, which is a key parameter characterizing the system’s behavior. This energy loss is defined as
\begin{align}
	E_s&=\hbar\omega_s I_s\nonumber\\
	&=\hbar\omega_s \frac{\gamma_e\gamma_c}{\gamma_e+\gamma_c}(\bar{n}_e-\bar{n}_c)
\end{align}
The energy released from the system to the environment at equilibrium is denoted by $E_s$.\\
 The energy the system gets from the emitter per unit of time, represented by $Q$, is then taken into account.  This amount can be expressed as 
\begin{equation}
	Q=f(\gamma_e,\gamma_c,T_e,T_c)\hbar\omega_s \bar{n}_e
\end{equation}
Here, $f(\gamma_e,\gamma_c,T_e,T_c)$ has dimensions of $time^{-1}$. The quantity $Q$ indicates the energy loss per unit time from the emitter due to the physical processes occurring within the system, and it is proportional to the energy of the fermions in the emitter reservoir. \\
The quantum transport factor $\eta_s$ offers important insights into the system's performance. It is defined as
\begin{align}
	\eta_s&= \frac{E_s}{Q}\nonumber\\
	&= \frac{1}{f}\frac{\gamma_e\gamma_c}{\gamma_e+\gamma_c} \left(1-\frac{\bar{n}_c}{\bar{n}_e}\right)\label{eff1}~.
\end{align}
The above expression's high-temperature limit will now be investigated.  From eq.(s) (\eqref{n}, \eqref{eff1}), we may formulate the high-temperature quantum transport factor as
\begin{align}
	\eta_s &= \frac{1}{f}\frac{\gamma_e\gamma_c}{\gamma_e+\gamma_c} \left(1-\frac{e^{\frac{\hbar\omega_s}{k_B T_e}}+1}{e^{\frac{\hbar\omega_s}{k_B T_c}}+1}\right)\nonumber\\
	&\approx \frac{1}{f}\frac{\gamma_e\gamma_c}{\gamma_e+\gamma_c} \left(1-\frac{2+\frac{\hbar\omega_s}{k_BT_e}}{2+\frac{\hbar\omega_s}{k_BT_c}}\right)\nonumber\\
	&\approx \frac{1}{f}\frac{\gamma_e\gamma_c}{\gamma_e+\gamma_c} \left[1-\left(1+\frac{\hbar\omega_s}{2k_BT_e}\right)\left(1-\frac{\hbar\omega_s}{2k_BT_c}\right)\right]\nonumber\\
    &=\frac{1}{f}\frac{\gamma_e\gamma_c}{\gamma_e+\gamma_c} \frac{\hbar\omega_s}{2k_BT_c} \eta_c\label{eff2}
\end{align}
where $\eta_{c}$ represents the well-known Carnot engine efficiency given by
\begin{align}
	\eta_c=1-\frac{T_c}{T_e}\label{carnot_effi}~.
\end{align}
Our next step is to determine the form of $f$. This can be accomplished by observing that the expression for $\eta_s$ must correspond to the well-known Carnot efficiency in the high-temperature limit. This condition directly leads to the conclusion that $f$ must equal $\frac{\gamma_e\gamma_c}{\gamma_e+\gamma_c}\frac{2k_BT_c}{\hbar\omega}$. As a result, the steady-state quantum transport factor can be expressed as
\begin{align}
	\eta_s&=\frac{2k_BT_c}{\hbar\omega}\left(1-\frac{\bar{n}_c}{\bar{n}_e}\right)\label{eff_steady}~.
\end{align}
This demonstrates unequivocally that the quantum transport factor has an upper constraint of unity.\\
Earlier in \cite{bhattacharya2025current}, it has been shown that the quantum transport factor for a bosonic system connected with two bosonic thermal reservoir becomes
\begin{align}
    \eta_b&=1-\frac{\bar{n}_c^b}{\bar{n}_e^b}\label{eff_boson}
\end{align}
where $\bar{n}_c^b$ and $\bar{n}_e^b$ denotes the bosonic thermal reservoir known as collector and emitter number density respectively. The expressions read
\begin{subequations}\label{n}
	\begin{align}
		\bar{n}_e^b&=\frac{1}{e^{\frac{\hbar\omega_s}{k_B T_e}}-1}\\
		\bar{n}_c^b&=\frac{1}{e^{\frac{\hbar\omega_s}{k_B T_c}}-1}~.
	\end{align}
\end{subequations}
Interestingly the bosonic quantum transport factor also matches Carnot efficiency given in eq. \eqref{carnot_effi} while temperature goes to infinity. But at low temperature the bosonic transport factor and fermionic transport factors are quite different in nature. The results are plotted in Fig. \ref{fig:fig1}. 
\section{Power spectrum of current}\label{sec3}
From eq. \eqref{current}, we found the current to be
\begin{align}
    I(t)&=\frac{1}{2}\left[\gamma_e\left({\bar{n}_e}-\left<\hat{n}\right>(t)\right)+\gamma_c\left(\left<\hat{n}\right>(t)-{\bar{n}_c}\right)\right]\nonumber\\
    &=\frac{1}{2}\left<\left(\gamma_e\bar{n}_e-\gamma_c\bar{n}_c\right)\hat{\mathbb{I}}-\left(\gamma_e-\gamma_c\right)\hat{a}^\dagger\hat{a}\right>\label{current_op1}
\end{align}
where $\hat{\mathbb{I}}$ is the identity operator and $\hat{n}=\hat{a}^\dagger\hat{a}$ is the fermionic number operator. Hence the current operator can be defined as
\begin{align}
    \hat{\iota}(t)=\frac{1}{2}\left[\left(\gamma_e\bar{n}_e-\gamma_c\bar{n}_c\right)\hat{\mathbb{I}}-\left(\gamma_e-\gamma_c\right)\hat{a}^\dagger\hat{a}\right]\label{current_op2}~.
\end{align}
The current is defined through the current operator as $I(t)=\left<\hat{\iota}(t)\right>$. Then we can write from eq. \eqref{current_op2}
\begin{align}
    \frac{d}{dt}\hat{\iota}&=-\left(\gamma_e+\gamma_c\right)\left(\hat{\iota}-I_s\hat{\mathbb{I}}\right)
\end{align}
where $I_s$ is is the steady current given by the eq. \eqref{is}.\\
Following the prescription presented in \cite{Carmichael} we can write the linear equations for currents in matrix form as
\begin{align}
    \frac{d}{dt} \begin{pmatrix}
        \hat{\iota}\\
        I_s\hat{\mathbb{I}}
    \end{pmatrix} &= \begin{pmatrix}
        -\gamma_e-\gamma_c & \gamma_e+\gamma_c\\
        0 & 0
    \end{pmatrix} \begin{pmatrix}
        \hat{\iota}\\
        I_s\hat{\mathbb{I}}
    \end{pmatrix}~.
\end{align}
Comparing the standard formulation we can identify 
\begin{align}
    \hat{\mathcal{A}}(t)=\begin{pmatrix}
        \hat{\iota}(t)\\
        I_s\hat{\mathbb{I}}
    \end{pmatrix}; \hspace{8mm} \mathcal{M}= \begin{pmatrix}
        -\gamma_e-\gamma_c & \gamma_e+\gamma_c\\
        0 & 0
    \end{pmatrix}\label{matrix1}
\end{align}
to write
\begin{align}
    \frac{d}{dt} \hat{\mathcal{A}}(t) &= \mathcal{M}\hat{\mathcal{A}}(t) \label{matrix}~.
\end{align}
For $\tau>0$, one gets from eq. \eqref{matrix} as
\begin{align}
    \frac{d}{d\tau} \left<\hat{\iota}(t)\hat{\mathcal{A}}(t+\tau)\right>&=\mathcal{M}\left<\hat{\iota}(t)\hat{\mathcal{A}}(t+\tau)\right>~.\label{fano1}
\end{align}
Putting the values from eq. \eqref{matrix1} in eq. \eqref{fano1}, we can write
\begin{align}
    \frac{d}{d\tau}\left<\hat{\iota}(t)\hat{\iota}(t+\tau)\right>&=\left(\gamma_e+\gamma_c\right)\left\{I_s I(t)-\left<\hat{\iota}(t)\hat{\iota}(t+\tau)\right>\right\}\label{fano2}~.
\end{align}
\begin{widetext}
Putting the value from eq. \eqref{current}, we can solve the differential equation with appropriate initial conditions. The solution reads
\begin{align}
    \left<\hat{\iota}(t)\hat{\iota}(t+\tau)\right>&=I_s^2+\left(\left<\hat{\iota}^2(t)\right>-I_s^2\right)e^{-\left(\gamma_e+\gamma_c\right)\tau}+\left(\gamma_e+\gamma_c\right)\left(I_s I_0-I_s^2\right)\tau e^{-\left(\gamma_e+\gamma_c\right)\tau}\label{fano3}~.
\end{align}
From eq. \eqref{matrix}, it can also be written for $\tau>0$ as
\begin{align}
    \frac{d}{d\tau} \left<\hat{\mathcal{A}}(t+\tau)\hat{\iota}(t)\right>&=\mathcal{M}\left<\hat{\mathcal{A}}(t+\tau)\hat{\iota}(t)\right>~.\label{fano4}
\end{align}
Then the current-current correlation can similarly be written as
\begin{align}
    \left<\hat{\iota}(t+\tau)\hat{\iota}(t)\right>&=I_s^2+\left(\left<\hat{\iota}^2(t)\right>-I_s^2\right)e^{-\left(\gamma_e+\gamma_c\right)\tau}+\left(\gamma_e+\gamma_c\right)\left(I_s I_0-I_s^2\right)\tau e^{-\left(\gamma_e+\gamma_c\right)\tau}\label{fano5}~.
\end{align}
Putting $t=0$ in eq. \eqref{fano3} and $t=-\tau$ in eq. \eqref{fano5}, we can get the current-current correlation for any $\tau$ as
\begin{align}
    \left<\hat{\iota}(0)\hat{\iota}(\tau)\right>&=I_s^2+\left(I_0^2-I_s^2\right)e^{-\left(\gamma_e+\gamma_c\right)|\tau|}+\left(\gamma_e+\gamma_c\right)\left(I_s I_0-I_s^2\right)|\tau| e^{-\left(\gamma_e+\gamma_c\right)|\tau|}\label{fano6}~.
\end{align}
From now on we can define the power spectrum \cite{huang2009introduction} from eq. \eqref{fano6} as
    \begin{align}
        S(\omega)&=\int_{-\infty}^{\infty} e^{i\omega\tau}\left<\hat{\iota}(0)\hat{\iota}(\tau)\right>d\tau\nonumber\\
        &=2\pi I_s^2\delta(\omega)+\frac{2\left(\gamma_e+\gamma_c\right)\left(I_0-I_s\right)}{\left(\gamma_e+\gamma_c\right)^2+\omega^2}\left\{I_0+I_s\frac{2\left(\gamma_e+\gamma_c\right)^2}{\left(\gamma_e+\gamma_c\right)^2+\omega^2}\right\}~.
    \end{align}
    \clearpage
\end{widetext}
The above relation is obtained through the integration
\begin{align}
    \int_{-\infty}^{\infty} e^{-\gamma|\tau|}e^{i\omega\tau}d\tau
    &=\int_{-\infty}^{0} e^{\gamma\tau}e^{i\omega\tau}d\tau+\int_{0}^{\infty} e^{-\gamma\tau}e^{i\omega\tau}d\tau\nonumber\\
    &=2\int_{0}^{\infty} e^{-\gamma\tau}\cos(\omega\tau)d\tau\nonumber\\
    &=\frac{2\gamma}{\gamma^2+\omega^2}~.
\end{align}
Similarly for the third term in eq. \eqref{fano6}, we have
\begin{align}
    \int_{-\infty}^{\infty}|\tau| e^{-\gamma|\tau|}e^{i\omega\tau}d\tau
    &=2\int_{0}^{\infty}\tau e^{-\gamma\tau}\cos(\omega\tau)d\tau\nonumber\\
    &=-\frac{\partial}{\partial\gamma}\left[\int_{-\infty}^{\infty} e^{-\gamma|\tau|}e^{i\omega\tau}d\tau\right]\nonumber\\
    &=2\frac{\gamma^2-\omega^2}{\left(\gamma^2+\omega^2\right)^2}
\end{align}
and 
\begin{align}
    \int_{-\infty}^{\infty} e^{i\omega\tau}d\tau&=2\pi\delta(\omega)
\end{align}
\begin{figure}[ht]
    \centering
    \includegraphics[width=0.9\linewidth]{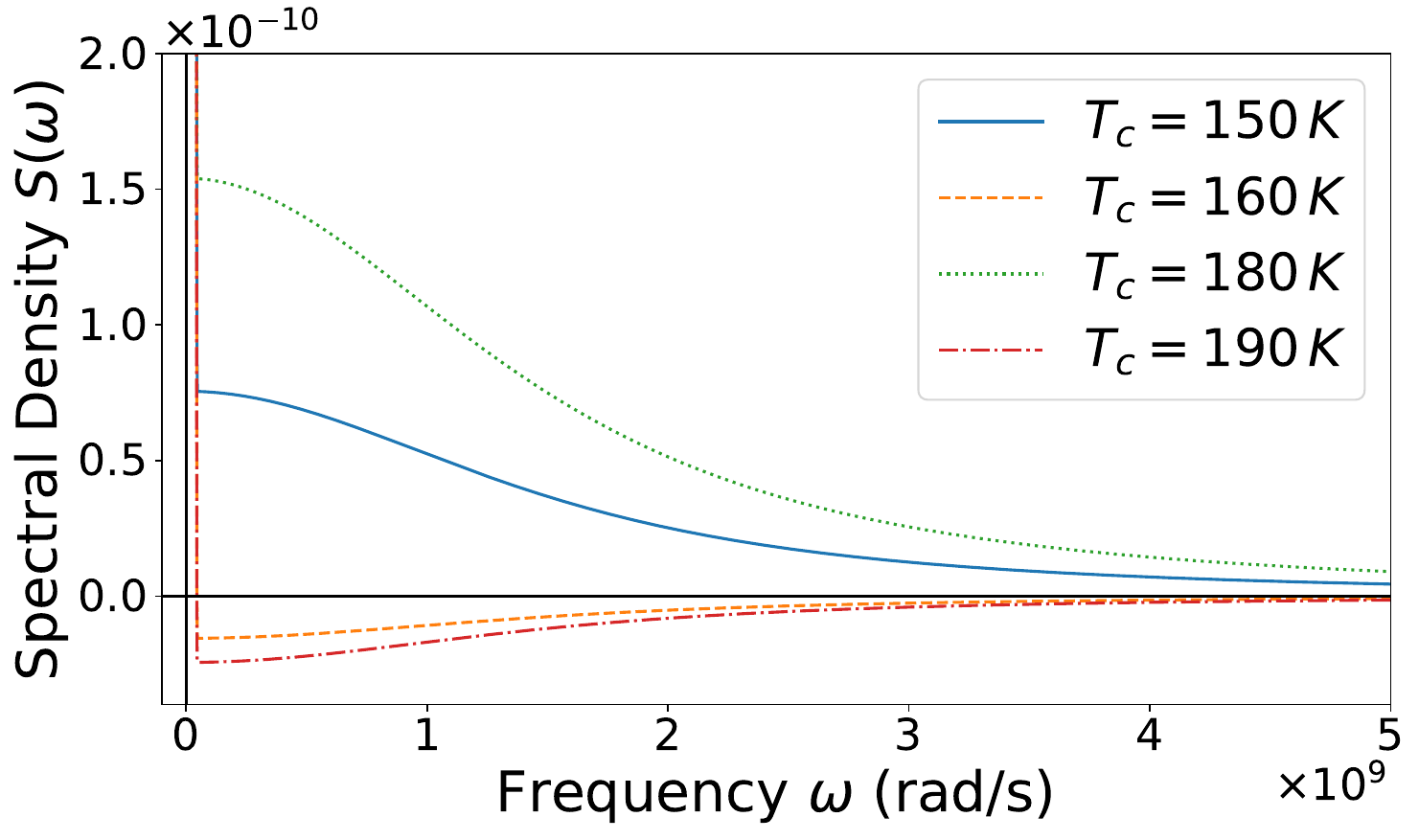}
    \caption{Spectral density is plotted against frequency at different collector temperatures while emitter temperature is kept constant at $300K$, system frequency is $10^{12}Hz$ and $\gamma_e=\gamma_c=10^9 Hz$.}
    \label{fig:fig2}
\end{figure}
\section{P-REPRESENTATION AND FOKKER-PLANCK EQUATION FOR Fermionic SYSTEM}\label{sec4}
\noindent In this section we aim to calculate the coherent state representation of the density matrix, known as Glauber-Sudarshan P-representation \cite{cahill1999density}. Fermionic creation and annihilation operators, unlike their bosonic counterparts, adhere to the following anticommutation relations, ensuring that the particles conform to Fermi-Dirac statistics
\begin{subequations}\label{grassmann1}
    \begin{align}
        &\anticom{ {a}_i}{ {a}^\dagger_j}=\delta_{ij}\\
        &\anticom{ {a}_i}{ {a}_j}=\anticom{ {a}^\dagger_i}{ {a}^\dagger_j}=0
    \end{align}
\end{subequations}
Fermions have only two possible energy eigenstates. Designating these as number states, the fermionic step-down and step-up operators act on them as follows
\begin{subequations}\label{grassmann2}
    \begin{align}
         {a}\ket{n}&=\sqrt{n}\ket{1-n}\\
         {a}^\dagger\ket{n}&=\sqrt{1-n}\ket{1-n}
    \end{align}
\end{subequations}
Given the unique nature of fermions, it is not possible to define a fermionic coherent state using ordinary numbers. Therefore, we adopt the prescription of \cite{das2019field,cahill1999density,ghosh2012fermionic} in this work. The normalized fermionic coherent state is thus defined as
\begin{align}
    |\xi\rangle &= D(\xi)|0\rangle \label{grassmann3}
\end{align}
Here, $D(\xi)$ represents the spin displacement operators, defined as
\begin{align}
    D(\xi) &= e^{ {a}^\dagger \xi-\xi^{*}  {a}}\nonumber\\
    &= 1+\left( {a}^\dagger \xi-\xi^{*}  {a}\right)+\left( {a}^\dagger  {a}-\frac{1}{2}\right) \xi^{*} \xi \label{grassmann4}
\end{align}
Clearly, $\xi$-s are not ordinary numbers. They are said to be the Grassmann numbers. For any mode $i$, $\xi_{i}$ and $\xi_{i}^{*}$ are Grassmann numbers for the corresponding mode, obeying the following anticommutation relations
\begin{subequations}\label{grassmann5}
\begin{align}
    \left\{\xi_{i}, \xi_{j}\right\} &= 0 \\
    \left\{\xi_{i}^{*}, \xi_{j}\right\} &= 0 \\
    \left\{\xi_{i}^{*}, \xi_{j}^{*}\right\} &= 0 ~.
\end{align}
\end{subequations}
We also assume the anticommutation of Grassmann numbers $\xi_{i}$ and $\xi_{i}^{*}$ with the operators $ {a}$ and $ {a}^\dagger$ as
\begin{subequations}\label{grassmann6}
\begin{align}
    \left\{\xi_{i}, a_{j}\right\} &= 0  \\
    \left\{\xi_{i}^{*}, a_{j}\right\} &= 0 \\
    \left\{\xi_{i}, a_{j}^{\dagger}\right\} &= 0 \\
    \left\{\xi_{i}^{*}, a_{j}^{\dagger}\right\} &= 0 ~.
\end{align}
\end{subequations}
Then the properties of displacement operator becomes evident. We can write
\begin{subequations}\label{grassmann7}
\begin{align}
     {D}^\dagger(\xi) {D}(\xi) &= \mathbb{I} \\
     {D}^\dagger(\xi)  {a}  {D}(\xi) &= ( {a} + \xi) \\
     {D}^\dagger(\xi)  {a}^\dagger  {D}(\xi) &= ( {a}^\dagger + \xi^*)
\end{align}
\end{subequations}
Now from eq. \eqref{grassmann3}, we can write the coherent state as
\begin{align}\label{coherent1}
    \ket{\xi}&=D(\xi)\ket{0}\nonumber\\
    &= \left[1+\left( {a}^\dagger \xi-\xi^{*}  {a}\right)+\left( {a}^\dagger  {a}-\frac{1}{2}\right) \xi^{*} \xi\right]\ket{0}\nonumber\\
    &=\left[1+{a}^\dagger \xi-\frac{1}{2} \xi^{*} \xi\right]\ket{0}~.
\end{align}
In second equation we used the identity $a\ket{0}=0$. Now we can write for coherent states
\begin{align}
    a\ket{\xi}&=aD(\xi)\ket{0}\nonumber\\
    &=D(\xi)D^\dagger(\xi)aD(\xi)\ket{0}\nonumber\\
    &=D(\xi)({a} + \xi)\ket{0}\nonumber\\
    &=\xi D(\xi)\ket{0}\nonumber\\
    &=\xi\ket{\xi}\label{coherent2}
\end{align}
and
\begin{align}
    a^\dagger\ket{\xi}&=a^\dagger\left[1+{a}^\dagger \xi-\frac{1}{2} \xi^{*} \xi\right]\ket{0}\nonumber\\
    &=\left[1-\frac{1}{2} \xi^{*} \xi\right] a^\dagger\ket{0}\nonumber\\
    &=\left[1-\frac{1}{2} \xi^{*} \xi\right] \left\{-\frac{\partial}{\partial\xi}\left( 1+a^\dagger\xi\right) \right\}\ket{0}\nonumber\\
    &=-\frac{\partial}{\partial\xi}\left(1+{a}^\dagger \xi-\frac{1}{2} \xi^{*} \xi\right)\ket{0}\nonumber\\
    &\hspace{8mm}+\frac{\xi^*}{2}\left(1+{a}^\dagger \xi-\frac{1}{2} \xi^{*} \xi\right)\ket{0}\nonumber\\
    &=\left(-\frac{\partial}{\partial\xi}+\frac{\xi^*}{2}\right)\ket{\xi}\label{coherent3}
\end{align}
where we used the identities like $(a^\dagger)^2=\xi^2=(\xi^*)^2=0$ in the above steps. Similarly we can write for the conjugate cases as
\begin{align}
    \bra{\xi}a^\dagger&=\xi^*\bra{\xi} \label{coherent4}\\
    \bra{\xi}a&=\left(\frac{\partial}{\partial\xi^*}+\frac{\xi}{2}\right)\bra{\xi} \label{coherent5}~.
\end{align}
Now like above expressions, we can rederive the results found in \cite{ghosh2012fermionic} as follows
\begin{subequations}\label{coherent6}
    \begin{align}
        a\ket{\xi}\bra{\xi}&=\xi\ket{\xi}\bra{\xi}\\
        a^\dagger\ket{\xi}\bra{\xi}&= \left(-\frac{\partial}{\partial\xi}+\xi^*\right)\ket{\xi}\bra{\xi}\\
        \ket{\xi}\bra{\xi}a^\dagger&= \xi^*\ket{\xi}\bra{\xi} \\
        \ket{\xi}\bra{\xi}a&=\left(\frac{\partial}{\partial\xi^*}+\xi\right)\ket{\xi}\bra{\xi} ~.
    \end{align}
\end{subequations}
These results will be necessary to calculate the Fokker-Planck equation from eq. \eqref{masteq}. From Glauber and Cahil's prescription,  we adopt the $P$-representation to express the density matrix as
\begin{align}
    \rho &=-\int d\xi^*d\xi P(\xi^*,\xi,t)\ket{\xi}\bra{\xi} ~. \label{P_rep}
\end{align}
Now we need to incorporate two rules for Grassmann variables before computing the Fokker-Planck equation. $\theta$ being a Grassmann variable, we can define the integration as
\begin{subequations}\label{grass_int}
    \begin{align}
        \int d\theta&=0\\
        \int \theta d\theta&=1
    \end{align}
\end{subequations}
and the Taylor series expansion of any function containing Grassman variable can be written as
\begin{align}
    f(\theta)&=f_0+f_1 \theta\label{taylor}
\end{align}
where $f_0$ and $f_1$ are the ordinary constants.\\
We can also define the Dirac delta function with Grassmanian variables. We know that the delta function has following property \cite{dirac1981principles}
\begin{align}
    \int d^2\xi \delta^{(2)}(\xi,\xi^*)=1~.
\end{align}
Then it is possible to define the delta function if and only if 
\begin{align}
    \delta^{(2)}~(\xi,\xi^*)&=\xi\xi^*~.\label{grassman_delta}
\end{align}
Since the probability distribution $P(\xi^*,\xi,t)$ must be an even function on both $\xi$ and $\xi^*$, it must be of the form
\begin{align}
    P(\xi^*,\xi,t)&=P_0(t)+P_1(t)\xi^*\xi\label{P-expand}~.
\end{align}
Now using the above expressions, we can write eq. \eqref{masteq} in the following form
\begin{align}
    \int d^2 \xi\frac{\partial P}{\partial t}\ket{\xi}\bra{\xi}&=\int d^2 \xi P [\mathrm{I}+\mathrm{II}]\ket{\xi}\bra{\xi}\label{fpeq1}
\end{align}
where
\begin{align}
    \mathrm{I}&=-\frac{\gamma_e+\gamma_c}{2}\left( \xi\frac{\partial}{\partial\xi} + \xi^*\frac{\partial}{\partial\xi^*}+4\xi^*\xi \right)\label{eq51}
\end{align}
and
\begin{align}
    \mathrm{II}&= \left(\gamma_e\bar{n}_e+\gamma_c\bar{n}_c \right)\left(\frac{\partial^2}{\partial\xi^* \partial\xi}-\xi\frac{\partial}{\partial\xi} - \xi^*\frac{\partial}{\partial\xi^*}-2\xi^*\xi \right)~.\label{eq52}
\end{align}
\begin{widetext}
    Putting the results of eq.(s) (\eqref{eq51},\eqref{eq52}) in eq.\eqref{fpeq1}, we get by comparing both sides
    \begin{align}
        \int d^2 \xi\frac{\partial P}{\partial t}\ket{\xi}\bra{\xi} = -\frac{\gamma_e+\gamma_c}{2} &\int d^2 \xi \left(\frac{\partial}{\partial\xi}(\xi P) + \frac{\partial}{\partial\xi^*}(\xi^*P) + 4\xi^*\xi P \right)\ket{\xi}\bra{\xi}\nonumber\\
         &+\left(\gamma_e\bar{n}_e+\gamma_c\bar{n}_c \right) \int d^2 \xi \left(\frac{\partial^2 P}{\partial\xi^* \partial\xi}-\frac{\partial}{\partial\xi}(\xi P) - \frac{\partial}{\partial\xi^*}(\xi^*P)-2\xi^*\xi P \right)\ket{\xi}\bra{\xi}~.\label{fpeq2}
    \end{align}
\end{widetext}
Now we can look upon each term in eq. \eqref{fpeq2}. For that we remember eq. \eqref{P-expand}, and then we get for the first term
\begin{align}
    &\int d^2 \xi \frac{\partial}{\partial\xi}(\xi P)\ket{\xi}\bra{\xi}\nonumber\\ &= \int d^2 \xi \frac{\partial}{\partial\xi}(\xi P_0(t))\ket{\xi}\bra{\xi}\nonumber\\ &= P_0(t)\int d^2 \xi \ket{\xi}\bra{\xi}\nonumber\\ &=P_0 (t)\label{part1}~.
\end{align}
The second term of the above equation reads
\begin{align}
    &\int d^2 \xi \frac{\partial}{\partial\xi^*}(\xi^* P)\ket{\xi}\bra{\xi}\nonumber\\  &= \int d^2 \xi \frac{\partial}{\partial\xi^*}(\xi^* P_0(t))\ket{\xi}\bra{\xi}\nonumber\\ &= P_0(t)\int d^2 \xi \ket{\xi}\bra{\xi}\nonumber\\ &=P_0 (t)\label{part2}~,
\end{align}
and the final term
\begin{align}
    &\int d^2 \xi \frac{\partial^2 P}{\partial\xi^* \partial\xi} \ket{\xi}\bra{\xi}\nonumber\\ &= \int d^2 \xi \frac{\partial^2 }{\partial\xi^* \partial\xi}\left( P_0(t) + P_1(t)\xi^*\xi \right) \ket{\xi}\bra{\xi} \nonumber\\ &= -P_1(t)~.
\end{align}
where we used the completeness relation
\begin{align}
    \int d^2 \xi \ket{\xi}\bra{\xi}&= 1~.
\end{align}
We use this completeness relation again to solve the last term in eq. \eqref{fpeq2}. We write
\begin{align}
    \int d^2 \xi~ \xi^*\xi P\ket{\xi}\bra{\xi}&= \int d^2 \xi ~\xi^*\xi P\int d^2 \chi \ket{\chi}\braket{\chi}{\xi}\bra{\xi}\label{eq61} ~.
\end{align}
The last integration over Grassmanian variable $\chi$ only survives if $\xi=\chi$. Then from eq. \eqref{eq61}, we get
\begin{align}
    \int d^2 \xi~ \xi^*\xi P\ket{\xi}\bra{\xi}&= \int d^2 \xi ~\xi^*\xi P\nonumber\\
    &= \int d^2 \xi ~\xi^*\xi \left( P_0(t)+P_1(t)\xi^*\xi \right)\nonumber\\
    &=-P_0~.
\end{align}
The differential equation then comes down from eq. \eqref{P-expand} to the form 
\begin{align}
    &\int d^2 \xi\frac{\partial P}{\partial t}\ket{\xi}\bra{\xi} =\alpha P_1(t) + \beta P_0(t)~.\label{eq64}
\end{align}
where $\alpha=\left(\gamma_e\bar{n}_e+\gamma_c\bar{n}_c \right)$ and $\beta = \gamma_e+\gamma_c$ are constants.\\
Now let us look upon the initial conditions and steady state conditions. In the steady state we know
\begin{align}
    \frac{\partial P}{\partial t}=0~.
\end{align}
Hence we get from eq. \eqref{eq64} 
\begin{align}
    \alpha P_1^{eq} + \beta P_0^{eq} = 0~.\label{eq67}
\end{align}
Again we can use the normalisation condition for the equilibrium as
\begin{align}
    \int d^2 &\xi P(\xi,\xi^*,t)=1\nonumber\\
    \int d^2 &\xi P_1(t)\xi^*\xi=1\nonumber\\
    &P_1(t) =-1~.\label{eq68}
\end{align}
Then at the steady state, comparing eq.(s) \eqref{eq67},\eqref{eq68}, we get
\begin{align}
    P_0^{eq} &=-\frac{\alpha}{\beta}=\frac{\gamma_e\bar{n}_e+\gamma_c\bar{n}_c}{\gamma_e+\gamma_c}\label{eq69}~.
\end{align}
Again, we can impose the initial condition as we did for the bosonic case. Initially, the distribution is considered located at a particular point, thus yielding a delta function. So the initial probability distribution will be
\begin{align}
    P^{in}(\xi,\xi^*,t=0)=\delta^{(2)}~(\xi,\xi^*)=\xi\xi^*.\label{eq70}
\end{align}
Thus the initial value for $P_0$ and $P_1$ are
\begin{align}
    P_0(0)&=0~;\label{eq71}\\
    P_1(0)&=-1~.\label{eq72}
\end{align}
Now let us chose a smooth differentiable function $f(t)$ such that,
\begin{align}
    P_0(t)&=\frac{\alpha}{\beta}f(t)\label{eq75}\\
    P_1(t)&=- 1\label{eq76}
\end{align}
From the boundary condition we can ensure that $f(0)=0$ and $f(\infty)=1$.\\
Also we can write the term
\begin{align}
    &\int d^2 \xi\frac{\partial P}{\partial t}\ket{\xi}\bra{\xi}\nonumber\\
    =&\int d^2 \xi\frac{\partial}{\partial t}\left(P_0(t)-\xi^*\xi\right)\ket{\xi}\bra{\xi}\nonumber\\
    =&\dot{P}_0(t)
\end{align}
where we used the facts that $\int d^2 \xi \ket{\xi}\bra{\xi}= 1$ and $P_1(t)=-1$. Hence the differential equation \eqref{eq64} becomes
\begin{align}
    \dot{P}_0(t) &= - \alpha  + \beta P_0(t)\nonumber\\
    \frac{\alpha}{\beta}\frac{df}{dt}&=-\alpha+\alpha f(t)\nonumber\\
    f(t)&=1-e^{-\beta t}~.\label{eqadd1}
\end{align}
Hence we have 
\begin{align}
    P_0(t)&=\frac{\alpha}{\beta}\left( 1-e^{-\beta t} \right)\label{eq78}\\
    P_1(t)&= -1~.\label{eq79}
\end{align}
Thus, the $P$ representation becomes
\begin{align}
    P(\xi,\xi^*,t)=\frac{\alpha}{\beta}\left( 1-e^{-\beta t} \right)- \xi^*\xi\label{eq80}~.
\end{align}
For a consistency check, we can find the steady state condition, that is $t\to\infty$, as
\begin{align}
    P(\xi,\xi^*,t\to\infty)&=\frac{\alpha}{\beta}-\xi^*\xi\nonumber\\
    &=\frac{\alpha}{\beta}e^{-\beta\xi^*\xi/\alpha}\label{eq81}~.
\end{align}
Clearly, the steady state distribution is a Gaussian distribution, as one finds for the bosonic case. 
\section{Conclusion} \label{sec5}
\noindent The dynamics of a fermionic oscillator connected to two fermionic reservoirs, $T_e$ and $T_c$, each with a different temperature, are examined in this article. With an emphasis on current flow, the quantum transport factor, and the power spectrum, we investigated a number of crucial facets of the system's behavior. We were able to characterize the development of the system and identify important relationships, including the current flow from the emitter to the collector, utilizing the flux balance condition by obtaining the master equation for the density operator under the Markovian approximation.\\
In order to evaluate the effectiveness of energy transmission within the system, we created the quantum transport factor. Our findings demonstrated that, despite certain parallels with classical efficiency measures, the quantum transport factor operates differently in fermionic systems. In particular, we found that, in contrast to the behavior seen in bosonic systems, the quantum transport factor can be less than the classical Carnot efficiency, especially at low temperatures. Our knowledge of quantum transport in fermionic systems is significantly expanded by this discovery.\\
Additionally, we conducted a comprehensive analysis of the power spectrum of the current by employing well-established techniques to compute the current-current correlation function. This approach enabled us to gain deeper insights into the frequency-dependent behavior of the current fluctuations and how they evolve with temperature. By systematically varying the temperature and examining the resulting changes in the power spectrum, we were able to identify key trends and underlying physical mechanisms governing the system's response. Our findings exhibited a strong qualitative agreement with previous studies, further reinforcing the validity of our methodology and confirming the robustness of our results. This consistency with prior work highlights the reliability of our approach and suggests that the techniques we employed can be effectively used to explore similar systems in future investigations.\\
Finally, in order to better comprehend the probability distribution of particles in the system, we extended our analysis by obtaining the Fokker-Planck equation. We were able to better manage the intricacy of fermionic states by using Grassmann algebra for fermionic operators, particularly when working with non-steady-state situations. Additionally, our work showed that steady-state results from previous research may be recovered by the time-dependent solution of the Fokker-Planck equation, bridging the gap between steady-state and non-steady-state analysis.\\
In conclusion, this work provides a thorough examination of fermionic transport dynamics, illuminating the complex behavior of quantum systems connected to fermionic reservoirs. Our results pave the way for further research in open quantum systems and their useful applications in quantum technologies by providing important new information on temperature, transport parameters, and quantum efficiency in fermionic systems.
\section*{Appendix}
\begin{figure}[ht]
    \centering
    \includegraphics[width=0.9\linewidth]{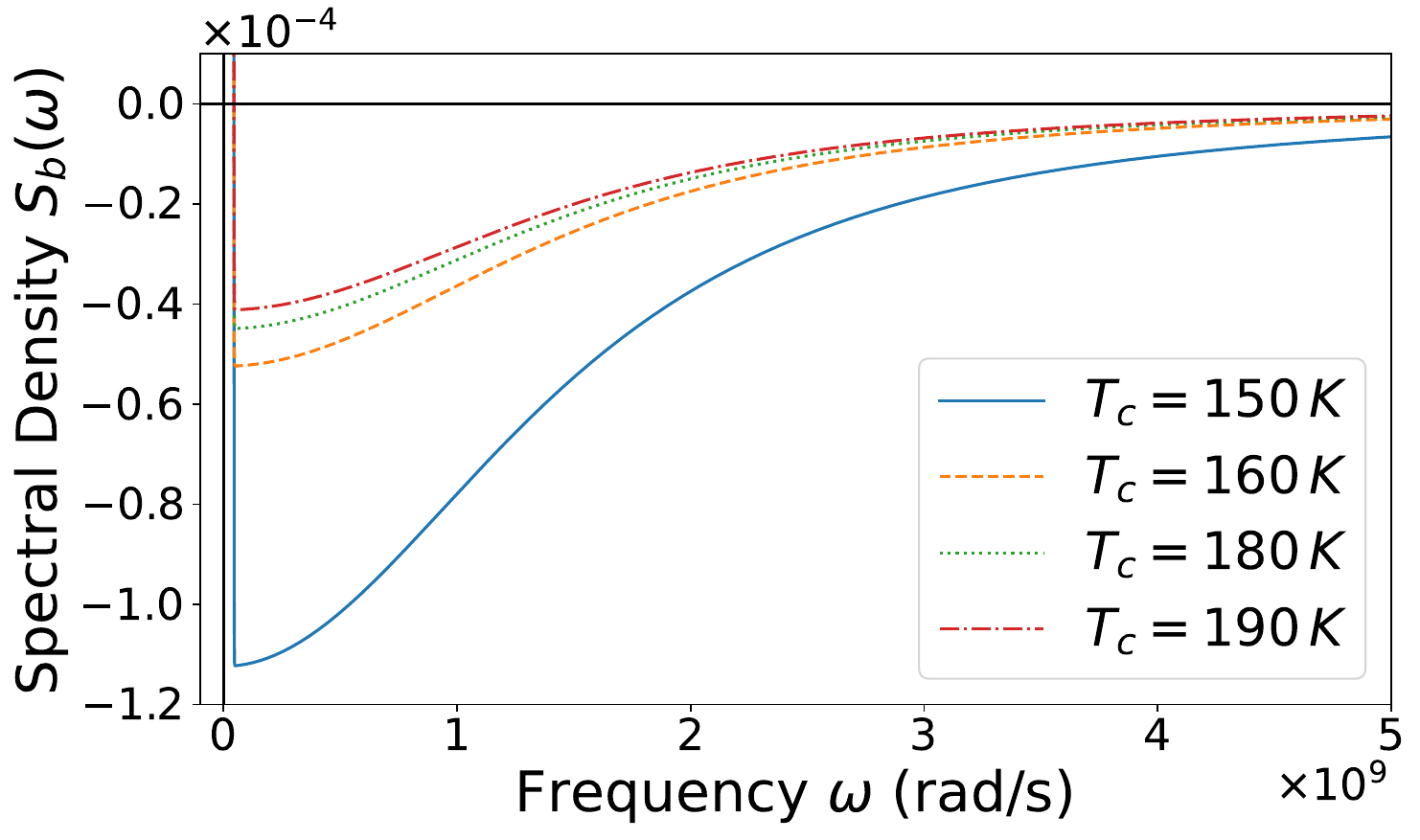}
    \caption{Spectral density for bosonic current is plotted against frequency at different collector temperatures while emitter temperature is kept constant at $300K$, system frequency is $10^{12}Hz$ and $\gamma_e=\gamma_c=10^9 Hz$.}
    \label{fig:fig3}
\end{figure}

In \cite{bhattacharya2025current}, we have calculated the current for a bosonic oscillator coupled with two bosonic thermal bath as
\begin{align}
    I_b(t)&=\frac{1}{2}\left[\gamma_e\left({\bar{n}^b_e}-\left<\hat{n}_b\right>(t)\right)+\gamma_c\left(\left<\hat{n}_b\right>(t)-{\bar{n}^b_c}\right)\right]\label{boson_current}
\end{align}
\begin{widetext}
where $\bar{n}_c^b$ are defined in eq.(s) \eqref{n} and $\hat{n}_b=\hat{b}^\dagger\hat{b}$. Here $\hat{b}$ and $\hat{b}^\dagger$ are the bosonic operators defined through the commutation relation
    \begin{align}
        &\comgen{\hat{b}}{\hat{b}^\dagger}=\mathbb{I}~;
        &\comgen{\hat{b}}{\hat{b}}=\comgen{\hat{b}^\dagger}{\hat{b}^\dagger}=0~.
    \end{align}
    The number density operator follows the relation, found in \cite{bhattacharya2025current} as
    \begin{align}
    	\left<\hat{n}_b\right>(t)&= \frac{\gamma_e\bar{n}^b_e+\gamma_c\bar{n}^b_c}{\gamma_e+\gamma_c} +\left(\left<\hat{n}_b(0)\right>-\frac{\gamma_e\bar{n}^b_e+\gamma_c\bar{n}^b_c}{\gamma_e+\gamma_c}\right)e^{-(\gamma_e+\gamma_c)t}\label{boson2}~.
    \end{align}
\end{widetext}
Eq. \eqref{boson_current} and eq. \eqref{boson2} structurally resembles with eq.(s) \eqref{current_op1} and \eqref{extra1} while the difference only appears in number distribution of boson and fermions. So, following the prescription described in \ref{sec3} we can calculate the current-current correlation for the boson as
\begin{widetext}
    \begin{align}
        S_b(\omega)&=2\pi (I^b_s)^2\delta(\omega)+\frac{2\left(\gamma_e+\gamma_c\right)\left(I^b_0-I^b_s\right)}{\left(\gamma_e+\gamma_c\right)^2+\omega^2}\left\{I^b_0+I^b_s\frac{2\left(\gamma_e+\gamma_c\right)^2}{\left(\gamma_e+\gamma_c\right)^2+\omega^2}\right\}
    \end{align}
\end{widetext}
where $I^b_s$ and $I^b_0$ are defined as
\begin{align}
    I^b_0&=\frac{1}{2}\left[\gamma_e\left({\bar{n}^b_e}-\left<\hat{n}_b\right>(0)\right)+\gamma_c\left(\left<\hat{n}_b\right>(0)-{\bar{n}^b_c}\right)\right]~;\\
	I^b_s&= \frac{\gamma_e\gamma_c}{\gamma_e+\gamma_c}(\bar{n}^b_e-\bar{n}^b_c)~.
\end{align}
The distribution can be visualised in Fig. \ref{fig:fig3}.

\bibliography{ref}

\end{document}